\documentstyle[aps,preprint]{revtex}

\def \grd {\mbox {\boldmath $\nabla$} }

\def \bsigma {\mbox {\boldmath $\sigma$} }

\begin{document}
\draft
\preprint{}
\title  {Spurious diffusion in particle simulations of the Kolmogorov flow}
\author{  M. Malek  Mansour $^{a}$, C. Van den Broeck $^{b}$, I.
Bena $^{b}$, and F. Baras $^{a}$}
\address{(a) Centre for
Nonlinear Phenomena and Complex Systems\\ Universit\'e Libre de Bruxelles,
Campus Plaine,
C.P. 231 \\ B-1050 Brussels, Belgium \\
(b)  Limburgs Universitair Centrum\\ B-3590 Diepenbeek, Belgium}
\date{\today}
\maketitle


\maketitle

\begin{abstract}
Particle simulations of the Kolmogorov flow are analyzed by the Landau-Lifshitz
fluctuating hydrodynamics.  It is shown that
a spurious diffusion of the center
of mass  corrupts the
statistical
properties of the flow. The analytical expression for the corresponding
diffusion coefficient is
derived.

\end{abstract}

Some
fifty years ago, Kolmogorov introduced a simple hydrodynamic model, that
displays various instabilities, including a transition to turbulence
\cite{obukhov}.
The richness of this model, known as the Kolmogorov flow,
combined with its simplicity, has attracted a great number of both theoretical
and numerical
work \cite{Sinai,Nepomniaschichii,Lorenz,She}.  In particular, particle
simulations, based on Lattice Boltzmann
\cite{LB} or
Lattice gas automata \cite{LG}, have been used to study the statistical
properties of the
flow in high Reynolds number regime \cite{She,henon,automata,BoonU}.  The
purpose of
this letter is to point out a subtle problem concerning the
nonequilibrium fluctuations that appear in this model. We show that
 the center of mass of the system undergoes a spurious diffusion that
corrupts the statistical properties of the flow.

The Kolmogorov flow is  an isothermal fluid confined in a rectangular box
$L_x \times L_y$, $\left\{ 0 \le x < L_x  ,  0  \le y < L_y \right\}$, with
periodic boundary
conditions in both directions. The flow is maintained through an external force
field of the form
\begin{equation}
{\bf   F}_{ext}=F_0\,\sin{(2\,\pi\,n\,y/L_y)}\,{\bf   1}_x,
\end{equation}
where ${\bf   1}_x$ is the unit vector in the $x$ direction.
The relevant parameters are the strength of the force field
$F_0$, the wave number $n$ of the forcing, and the aspect ratio
$a_r = L_x/L_y$.

For small enough $F_0$, the flow
follows basically the external field and the stationary velocity profile
is readily found to be
\begin{equation}
\label{v}
{\bf v}_{st}  =  u_0 \,\sin{(2\,\pi n \, y/L_y)}\,{\bf 1}_x \, ,
\end{equation}
\begin{equation}
\label{u0}
 u_0  \equiv  \frac{F_0\,L_y^2}{4\,\pi^2\,n^2 \, \eta}
\end{equation}
where $\eta$ is the shear viscosity coefficient.  However, upon increasing
$F_0$, this
stationary state becomes unstable giving rise to rotating convective patterns
\cite{Sinai,Nepomniaschichii}. Other instabilities of increasing complexity
appear for larger values of $F_0$, culminating in a turbulent - like behavior
\cite{Lorenz,She}.

To study the statistical properties of this system, we turn to a description
in terms of the fluctuating hydrodynamic equations:
\begin{equation}
\frac{\partial \,\rho}{\partial\,t} \, = \, -\grd\,\cdot\,(\rho\,{\bf   v})
\label{continuity}
\end{equation}
\begin{equation}
\frac{\partial\,(\rho\,{\bf   v})}{\partial\,t} \, =\, -\grd\,\cdot\,(\rho
\,{\bf
v} \,{\bf  v}) \,-\,\grd\,P \,- \, \grd \cdot \bsigma \,+\, {\bf   F}_{ext},
\label{momentum}
\end{equation} where $\rho$ is the mass density, $P$ the hydrostatic
pressure and
$\bsigma$ the {\it two-dimensional} stress tensor:
\begin{equation}
\sigma_{i, j} \, =\, - \, \eta \, \left(
\frac{\partial\,v_i}{\partial\,x_j} + \,
\frac{\partial\,v_j}{\partial\,x_i} - \, \delta_{i, j} \, \grd \cdot {\bf
v} \right) \, - \,
\zeta \, \delta_{i, j} \, \grd \cdot {\bf   v} \, +  S_{i, j}.
\label{stress}
\end{equation}
${\bf   S}$ is a random tensor whose elements $\{ S_{i, j} \}$ are
Gaussian white noises with zero mean and covariances given by \cite{Landau}
\begin{eqnarray}
< S_{i, j}({\bf   r}, t) \, S_{k, \ell}({\bf   r}', t') > & = & 2 k_B T_0 \,
\delta (t - t')  \, \delta ({\bf   r} -{\bf   r}') \nonumber\\
& &\left [ \eta (\delta_{i,
k}^{Kr}
\delta_{j, \ell}^{Kr} +
\delta_{i,\ell}^{Kr} \delta_{j, k}^{Kr} ) \, + \, (\zeta - \eta) \delta_{i,
j}^{Kr}
\delta_{k, \ell}^{Kr} \right ]
\label{noisecor}
\end{eqnarray}
where $k_B$ and $T_0$ stand for the Boltzmann constant and the (uniform)
temperature,
respectively.  For simplicity, we shall assume that the shear and bulk
viscosity
coefficients,
$\eta$ and $\zeta$, are  state independent, i.e., they are constant.

When imposing a force field, one has to keep in mind
that both in microscopic
simulations and in real systems, the fluid is made out of individual
particles. Hence one cannot impose a bulk force, but rather an acceleration
field acting on the  particles.  Since the density of particles is
fluctuating, we conclude that the external field in the momentum
equation (\ref{momentum}) is also a fluctuating quantity:
\begin{equation}\label{FF}
{\bf F}_{ext} = \rho(x, y) \, a_0 \, \sin{(2\,\pi\,n\,y/L_y)}\,{\bf 1}_x,
\end{equation}
where $a_0$ is the amplitude of the imposed acceleration field.
Furthermore, since the external field in the Kolmogorov flow is
space-dependent, the force acting on a particle
depends on its {\it exact} position so that the total force (in the
$x$-direction)
$F(t)$ will also be fluctuating, even though the total number of particles is
conserved. As a result, the center of mass linear momentum, denoted by
$J_x(t)$, undergoes a stochastic motion driven by a scalar force $F(t)$:
\begin{equation}
\frac{\partial J_x(t)}{\partial t} \,=\, \frac{a_0}{L_x\;L_y} \,
\int_{0}^{L_x}\,dx \int_{0}^{L_y} \,dy \, \rho(x,y, t)
\,\sin{(2\,\pi\,n\,y/L_y)} \, \equiv \, F(t).
\label{cm}
\end{equation}

In strictly subsonic regimes the flow behaves essentially as an
incompressible fluid so
that the average density is uniform in space, $<\rho> = \rho_0$.  It then
follows from eq.
(\ref{cm}) that $<F(t)> = 0$.  To find the force correlation function
$<F(t) F(t')>$, we consider
the spatial average of the hydrodynamical equations (\ref{continuity},
\ref{momentum}) over
the $x$ direction and notice that corresponding spatially averaged density,
$\rho(y,t)=\frac{1}{L_x} \,
\int_{0}^{L_x}\,dx \, \rho(x,y,t)$, and
$y$ component of the
velocity, $v(y,t)=\frac{1}{L_x} \,
\int_{0}^{L_x}\,dx \, v_y(x,y, t)$, are not affected by the external
constraints, i.e. they assume their
equilibrium form.  In particular, in the {\it stationary
regime} one has that
$<\rho> =
\rho_0$ and $<v> = 0$ are
independent of the value of $a_0$.  To study the fluctuations around this
state, we introduce the deviations
$\delta \rho(y,
t)= \rho(y,t)\,-\,\rho_0$, $\delta v(y, t)=v(y, t)
$ and $\delta P(y, t) = P(y, t)\,-\,<P>$, that obey the following
linearized equations :
\begin{eqnarray}
\frac{\partial\,\delta\rho}{\partial t} & = & - \, \rho_0 \,
\frac{\partial\,\delta
v}{\partial y} \label{er}\\
\rho_0 \, \frac{\partial\,\delta v}{\partial\,t} & = & -\,\frac{\partial \,
\delta P}{\partial
\, y} \, + \, (\eta + \zeta) \, \frac{\partial^2 \, \delta v}{\partial \, y^
2}
\, -\frac{\partial \, S_{yy}}{\partial \, y} \label{ev}
\end{eqnarray}
with
\begin{equation}
< S_{yy}(y, t) \,S_{yy}(y', t') >\, = \, 2 \frac{ k_B T_0}{L_x} (\eta +
\zeta) \,
\delta (t - t')  \, \delta (y -y')
\end{equation}
To close these equations, we need to specify the equation of state.  Since
the fluid is isothermal, we simply set
\begin{equation}
\delta P\,=\,c_s^2\,\delta\rho\,,
\end{equation}
where $c_s$ is the isothermal sound speed.

 The
 stochastic differential evolution equation for the fluctuating force
$F(t)$ now follows  easily  by multiplication of (\ref{er}) and
(\ref{ev}) by
$\sin{(2\,\pi\,n\,y/L_y)}$ and
$\cos{(2\,\pi\,n\,y/L_y)}$, respectively, followed by integration over $y$. One
obtains:
\begin{equation} \label{stoceq}
\frac{d^2\,F(t)}{dt^2}\,+\,\frac{\eta\,+\,\zeta}{\rho_0}\,\frac{4 \pi^2
n^2}{L_y^2}\,\frac{d\,F(t)}{dt}\,+\,\frac{4 \pi^2 n^2
c_s^2}{L_y^2}\,F(t)\,=\,\psi(t)\,,
\end{equation}
 where $\psi(t)$ is a white Gaussian noise, with zero mean and
variance given by:
\begin{equation}\label{ef}
< \psi (t)\,\psi (t') > \,= \,\frac{k_B\,T_0}{L_x L_y}\,(\eta\,+\,\zeta) \,
a_0^2 \, \bigg(
\frac{2 \pi \, n}{L_y} \bigg)^4\,\delta(t\,-\,t')\,.
\end{equation}
We conclude that $F(t)$ is a Gaussian
non-Markovian process.
The exact form of the
force correlation is easily obtained from (\ref{stoceq}) and (\ref{ef}),
but the final
expression is rather lengthy. On the other hand, the validity of
hydrodynamics can only be
guaranteed if the parameter
\begin{equation}
\epsilon\,= \,\frac{\eta + \zeta}{\rho_0 \, c_s\,L_y}
\end{equation}
remains small \cite{Alder}. Accordingly, to dominant order in $\epsilon$,
the force
correlation reads:
\begin{equation}
<F(t) F(0)> \, = \, \rho_0^2 \, a_0^2 \, \, \frac{k_B \, T_0}{2 \, m \, N \,
c_s^2} \, \exp
\left\{ - \, 4 \pi^2 n^2 \, \Gamma_s \, t / L_y^2 \right\} \cos( 2 \pi n\,
c_s \, t /
L_y) \, \, \, \, ; \, \, \,\, t \, \ge \, 0,
\end{equation}
where $\Gamma_s = (\eta + \zeta)/ 2 \rho_0$ represents the (two
dimensional) sound
damping coefficient, $N$ is the total number of particles and  $m$ their
individual mass.

Turning to $J_x(t)$, which is nothing but the time integral of $F(t)$, we
conclude that it is a Gaussian  stochastic process with zero average
and second moment (again to dominant order in $\epsilon$):
\begin{eqnarray}
\label{jx2}
<J_x^2(t)> & = & \rho_0^2 \, a_o^2 \, \, \frac{k_B \, T_0}{ m \, N \,
c_s^4} \,\bigg( \, \Gamma_s
\, t
\nonumber\\  & + & \, \left( 2 \pi n / L_y \right) ^{-2} \left[\, 1 \, - \,
\exp
( - \, 4 \pi^2 n^2 \, \Gamma_s \, t / L_y^2 ) \,\, \cos( 2 \pi n\, c_s \, t /
L_y) \right] \, \bigg)
\end{eqnarray}

As announced, $J_x(t)$ diffuses in time (in the momentum space) with a
long-time diffusion
coefficient given by:
\begin{equation}
\label{diff}
D \, = \, \lim\limits_{t \rightarrow \infty} \, \, \frac{<J_x^2(t)>}{t} \,  =
\,
\rho_0^2 \, a_0^2
\, \, \frac{k_B \, T_0}{ m \, N \, c_s^4} \,\, \Gamma_s.
\end{equation}

It is important to notice that in real macroscopic systems the very
existence of the center of
mass diffusion remains questionable for the following reason. Periodic
boundary conditions are
one of the basic simplifying features of the Kolmogorov flow.  This is fine
for the
representation of a system of infinite extend, consisting of periodically
repeated Kolmogorov
units, as long as only macroscopic properties are at stake. However, when
the fluctuations are
under investigation, one has to realize that the periodic boundary
conditions imply a perfect
correlation of the fluctuating forces in the different units. This is
obviously unphysical
(except for the academic situation of a system defined on a torus). In any
case, the
diffusion coefficient $D$ is negligibly small in macroscopic systems
(typically of the order
of $10^{-38} kg^2/m^2 s^3$) and is  essentially unobservable.

The situation is entirely different in microscopic simulations where the
total number of particles $N$ barely exceeds $10^5$. To estimate the
importance of the center of mass
diffusion, and the corresponding contamination of statistical properties of
the system in numerical
simulations, we first note that the ratio $k_B\,T_0 / m \, c_s^2$ is of the
order of
unity.  We next observe that there is a minimum run
time for simulations, namely the hydrodynamic relaxation time
$\tau_h \approx L_x L_y/\Gamma_s$. Typical running times are several such
$\tau _h$. It then follows from eq. (\ref{jx2}) that for large $t$  (i.e.  $t >
\tau_h$) the center of mass velocity fluctuation, $<J_x^2>/\rho_0^2$, is about:
\begin{equation}
\label{vx2}
 <v_x^2(t)> \, = \, \frac{<J_x^2(t)>}{\rho_0^2} \,  \approx
 \, \frac{a_0^2}{ n_0 \, c_s^2} \,\, \frac{t}{\tau_h},
\end{equation}
where $n_0 = N/L_x L_y$ is the number density.  This quantity has to be
compared with the spatial average of the mean square
flow velocity ${\bar u_m^2}$, which is of the order of $u_0^2/2$  (see eq.
(\ref{u0})).  The relative
importance of the center of mass diffusion can thus be estimated by the
square root of the ratio
$<v_x^2(t)> / {\bar u_m^2}$, hereafter denoted by $\mu(t)$. Using the
explicit form of $u_0$, eq.
(\ref{u0}), one finds that for $t > \tau_h$,
\begin{equation}
\label{mu}
\mu(t) \, = \, \big(<v_x^2(t)> / \, {\bar u_m^2}\, \big)^{1/2} \,
\approx \,  \big( \, 2 / n_0 \big)^{1/2} \, a_r \,\,
\frac{4 \pi^2 \,n^2 \, \eta }{m \, N \, c_s} \,\, \big( t / \tau_h \big)^{1/2}
\end{equation}
where $a_r = L_x/L_y$ is the aspect ratio.

As a first example, we consider a two dimensional Boltzmann gas for
which there exists an efficient algorithm, proposed two decades ago by Bird,
that is about 3 orders of magnitude faster than the corresponding traditional
molecular dynamic simulation \cite{Bird}.  A  typical case is a
system involving
$20\,000$ hard disks of diameter $d$, with $L_x \times L_y$ $=2000
\times 1000  \, d^2$ (i.e. $a_r = 2$ and $n_0 = 10^{-2}$ particles per $d^2$),
$n = 2$,  $c_s \approx 1$ and $\eta \approx 0.3$ (in system units, where
lengths, masses and velocities are
scaled by the disk diameter $d$, the particle mass $m$ and the thermal
velocity, $\sqrt{k_B T_0 /
m}$, respectively). It then follows from eq. (\ref{mu}) that after only one
relaxation time,
 $\mu(\tau_h) \approx 7 \times 10^{-2}$ which is certainly not negligible,
all the more so since
typical running times are 10 to 100 times larger than $\tau_h$.

One way to avoid this problem is to increase the number of particles, while
keeping the number density $n_0
= 10^{-2}$ particles per $d^2$,  since then the Bird algorithm is
applicable. However, to reach
reasonably small values of
$\mu$, like for instance $\mu(\tau_h) \approx 10^{-4}$, one has to consider
simulations involving over $13$
millions of particles.  Such simulations require a prohibitively long
running time with
present day computers.

The only other alternative is to increase the number density as well. For a
given number of particles, the
best strategy is to choose $n_0$ so that the Reynolds number is as high as
possible, since this is
precisely one of the main objectives of numerical simulations
\cite{adjustable}.  In the case of subsonic hard disk flows, the
appropriate number density turns out to
be about $n_0 = 0.27$ particles per
$d^{2}$
\cite{Puhl}.  For a system containing half a million of particles,  $L_x
\times L_y$ $=960 \times 1920
\, d^2$, $c_s
\approx 1.6$ and $\eta \approx 0.4$.  The function $\mu(\tau_h)$ is then
about $4 \times 10^{-4}$,
which is quite satisfactory.   However, a number density of $n_0 \approx
0.27$ corresponds to a moderately dense Enskog gas for which the Bird
algorithm is no longer applicable
\cite{densebird}. Instead, one has to use the traditional hard disk
molecular dynamics method which, as
mentioned before, is about 3 orders of magnitude slower than the
corresponding dilute gas simulation.
Furthermore, the collision frequency grows linearly with the number
density, which further increases the
running time by at least another order of magnitude. Under these
conditions, pursuing the
simulation for a single relaxation time
$\tau_h$ is about the best one can achieve with present day computer
performances.  Although such a
relatively short simulation might be satisfactory to study the average
properties of the
system, it is certainly not enough to extract the associated fluctuation
spectrum.

The above discussion highlights the usefulness of lattice-particle simulations
for the study of the relatively high Reynolds number flows.  But these
model simulations have their own limitations.  Because the motion of
particles takes place within a restricted geometry (4 or 6 linear directions),
with the corresponding restricted number of velocities, reaching local
equilibrium requires now many more collisions than in the case of hard disk
dynamics \cite{rothman}.  As
far as macroscopic properties of the system are concerned, this is only a
minor problem, since
lattice-particle simulations  typically run seven orders of magnitude faster
than hard disk molecular dynamics. The major drawback however is that
such a long time simulation inevitably increases the effect of the center
of mass diffusion
reported here. In fact, the spurious diffusion  has  been noticed very
recently by Boon et al. \cite{boon} in a  study of the so-called "turbulent
diffusion" in Kolmogorov flow.

In conclusion, while spurious diffusion of the center of mass in the
Kolmogorov flow does not affect the average macroscopic behavior of the
system, it does corrupt the other statistical properties, and to a
significant degree under conditions that are typical for many microscopic
simulations. The best way to avoid this problem is to include in the
simulation
algorithm an ad-hoc mechanism that prevents the center of mass momentum
fluctuations.  This can be accomplished rather easily in
lattice-particle simulations \cite{Hanon}, but
its counterpart in molecular dynamic simulations is less obvious.

\section{Acknowledgments}
{We thank G. Nicolis and
J. W. Turner for stimulating discussions. We acknowledge  support from the
IUAP,
Prime Minister's Office of  the Belgian
Government, and one of us (CVdB) also from the FWO Vlaanderen.  This work
has been partly supported by the
European Commission DG 12 (Advanced Research Meetings and Studies).}

\end{document}